\documentclass[a4paper,11pt]{article}
\usepackage{jinstpub} % for details on the use of the package, please see the JINST-author-manual
\usepackage{lineno}
\title{\boldmath Status of the detector setup for the FAMU experiment at RIKEN-RAL for a precision measurement of the Zemach radius of the proton in muonic hydrogen}

\author[a,b,c,*]{R. Rossini,\note[*]{Corresponding author.}}
\author[e]{A. Adamczak,}
\author[f]{D. Bakalov,}
\author[g,h]{G. Baldazzi,}
\author[i]{S. Banfi,}
\author[d]{M. Baruzzo,}
\author[i]{R. Benocci,}
\author[i]{R. Bertoni,}
\author[i,j]{M. Bonesini,}
\author[d]{V. Bonvicini,}
\author[d]{H. Cabrera,}
\author[i,k]{S. Carsi,}
\author[d,l]{D. Cirrincione,}
\author[i,j]{M. Clemenza,}
\author[m,n]{L. Colace,}
\author[o]{M. B. Danailov,}
\author[f]{P. Danev,}
\author[a,b]{A. de Bari,}
\author[b]{C. de Vecchi,}
\author[p,q]{E. Fasci,}
\author[d,r,s]{K. S. Gadedjisso-Tossou,}
\author[i]{R. Gaigher,}
\author[p,q]{L. Gianfrani,}
\author[c]{A. D. Hillier,}
\author[c,t]{K. Ishida,}
\author[c]{P. J. C. King,}
\author[i]{V. Maggi,}
\author[a,b]{A. Menegolli,}
\author[d]{E. Mocchiutti,}
\author[p,q]{L. Moretti,}
\author[g,h,u]{G. Morgante,}
\author[r]{J. Niemela,}
\author[i,k]{C. Petroselli,}
\author[v]{A. Pirri,}
\author[d,]{C. Pizzolotto,}
\author[b]{M.C. Prata,}
\author[v,w,x]{A. Pullia,}
\author[b,3]{M. Pullia,}
\author[x,y]{R. Ramponi,}
\author[b]{M. Rossella,}
\author[1]{R. Sarkar,}
\author[h]{A. Sbrizzi,}
\author[f]{M. Stoilov,}
\author[d]{J. J. Suarez-Vargas,}
\author[2]{G. Toci,}
\author[m]{L. Tortora,}
\author[i,k]{E. S. Vallazza,}
\author[c]{K. Yokoyama,}
\author[d,l,t]{A. Vacchi}

\affiliation[a]{Dept. of Physics, University of Pavia, Pavia, Italy}
\affiliation[b]{Sezione di Pavia, Istituto Nazionale di Fisica Nucleare (INFN), Pavia, Italy}
\affiliation[c]{ISIS Neutron and Muon Source, Science and Technology Facilities Council (STFC), Didcot, UK}
\affiliation[d]{Sezione di Trieste, Istituto Nazionale di Fisica Nucleare (INFN), Trieste, Italy}
\affiliation[e]{Institute of Nuclear Physics, Polish Academy of Sciences, Kraków, Poland}
\affiliation[f]{Institute for Nuclear Research and Nuclear Energy, Bulgarian Academy of Sciences, Sofia, Bulgaria}
\affiliation[g]{Dept. of Physics "A. Righi", University of Bologna, Bologna, Italy}
\affiliation[h]{Sezione di Bologna, Istituto Nazionale di Fisica Nucleare (INFN), Bologna, Italy}
\affiliation[i]{Sezione di Milano-Bicocca, Istituto Nazionale di Fisica Nucleare (INFN), Milan, Italy}
\affiliation[j]{Dept. of Physics "G. Occhialini", University of Milano-Bicocca, Milan, Italy}
\affiliation[k]{Dept. of Science and High Technology, University of Insubria, Como, Italy}
\affiliation[l]{Dept. of Mathematics, Computer Science and Physics, University of Udine, Udine, Italy}
\affiliation[m]{Sezione di Roma Tre, Istituto Nazionale di Fisica Nucleare (INFN), Rome, Italy}
\affiliation[n]{Dept. of Engineering, University of Roma Tre, Rome, Italy}
\affiliation[o]{Elettra Sincrotrone Trieste, Basovizza (Trieste), Italy}
\affiliation[p]{Dept. of Mathematics and Physics, University of Campania Luigi Vanvitelli, Caserta, Italy}
\affiliation[q]{Sezione di Napoli, Istituto Nazionale di Fisica Nucleare (INFN), Naples, Italy}
\affiliation[r]{The "Abdus Salam" International Centre for Theoretical Physics (ICTP), Trieste, Italy}
\affiliation[s]{Dept. of Physics, University of Lomè, Lomè, Togo}
\affiliation[t]{RIKEN Nishina Center, Saitama, Japan}
\affiliation[u]{Osservatorio di Astrofisica e Scienza dello Spazio (OAF), Istituto Nazionale di Astrofisica (INAF), Bologna, Italy}
\affiliation[v]{Istituto di Fisica Applicata “N. Carrara” (IFAC), Consiglio Nazionale delle Ricerche (CNR), Sesto Fiorentino, Italy}
\affiliation[w]{Dept. of Pysics "A. Pontremoli", University of Milan, Milan, Italy}
\affiliation[x]{Sezione di Milano, Istituto Nazionale di Fisica Nucleare (INFN), Milan, Italy}
\affiliation[y]{Istituto di Fotonica e Nanotecnologie (IFN), Consiglio Nazionale delle Ricerche (CNR), Milano, Italy}
\affiliation[z]{Institute of Solid Physics, Bulgarian Academy of Sciences, Sofia, Bulgaria}
\affiliation[1]{Indian Centre for Space Physics, Kolkata, India}
\affiliation[2]{Istituto Nazionale di Ottica (INO), Consiglio Nazionale delle Ricerche (CNR), Sesto Fiorentino, Italy}
\affiliation[3]{Centro Nazionale di Adroterapia Oncologica (CNAO), Pavia, Italy}
\emailAdd{riccardo.rossini@infn.it}

\abstract{The FAMU experiment at RIKEN-RAL is a muonic atom experiment with the aim to determine the Zemach radius of the proton by measuring the 1s hyperfine splitting in muonic hydrogen. The activity of the FAMU Collaboration in the years 2015-2023 enabled the final optimisation of the detector-target setup as well as the gas working condition in terms of temperature, pressure and gas mixture composition. The experiment has started its data taking in July 2023. The status of the detector setup for the 2023 experimental runs, for the beam characterisation and muonic X-ray detection in the 100-200 keV energy range, is presented and discussed.}

\begin{document}
\maketitle
\flushbottom

\section{Introduction}\label{sec:intro}
    The first measurements of the 2s-2p Lamb shift in muonic hydrogen ($\mu p$) returned a value of proton charge radius equal to $0.84087(39)$ fm\cite{antognini2013}, 7$\sigma$ away from the accepted value from CODATA 2010 $0.8775(51)$ fm\cite{carlson2015}, which gave rise to the so-called \emph{proton radius puzzle}. Many efforts are being made in order to shed light on this puzzle, also in the field of \emph{ep} scattering\cite{xiong2019}.
    
    The FAMU (\emph{Fisica degli Atomi MUonici} - Muonic Atom Physics) experiment aims to measure the proton Zemach radius ($r_Z$) with an experimental uncertainty in the order of 1\%\cite{pizzolotto2020}. The experiment consists in measuring the hyperfine splitting \emph{hfs} in 1s muonic hydrogen, which is strongly dependent on $r_Z$\cite{carlson2015, pizzolotto2020, carlson2011, faustov2014, dupays2003, volotka2005}, with an accuracy better than $10^{-5}$. As of today, this value has been extracted from \emph{hfs} in ordinary hyrogen $ep$\cite{dupays2003, volotka2005, brodsky2005} as well as from the 2s \emph{hfs} in $\mu p$\cite{antognini2013}, but not from $\mu p$ \emph{hfs}.

    The FAMU experimental methods consists in forming $\mu p$ atoms using a pulsed low-momentum $\mu^-$ beam at the RIKEN-RAL Port1 beamline\cite{matsuzaki2001, hillier2019}, and looking for a specific X-ray signature of the excitation of the \emph{hfs} as a function of the wavelength from a tunable Mid-InfraRed (MIR) laser. This signature consists of an excess of delayed muonic hydrogen X-rays in the 100-200 keV region, as explained in Section \ref{sec:target}. The FAMU experimental method yields some strict requirements for the experiment detectors. 

    The presence of a muon beam hodoscope in the FAMU setup is important for many reasons. First, the beam shape had to be optimised in order to maximise the amount of muons entering the hydrogen target. Moreover, the amount of delayed oxygen X-rays depends on the number of $\mu p$ atoms in the chamber, and therefore on the muon flux in each muon spill. As this value depends on the ISIS synchrotron current, which can face small variations in time, muon beam monitoring is crucial for data normalisation. The FAMU beam hodoscope was calibrated at a low-rate proton beam at the CNAO synchrotron in Pavia, Italy. Further details on the beam hodoscope are presented in Section \ref{sec:hodo}.

    Now focusing on the X-ray detectors, the FAMU experimental technique requires a good compromise among time resolution performance, efficiency and energy resolution at 100-200 keV. As a consequence, it was decided to focus on inorganic scintillating crystals. %In particular, all FAMU scintillators are based on lanthanum bromide scintillating crystals (LaBr$_3$:Ce). 
    The optimisation of timing performances first led to read out the scintillation light using PhotoMultiplier Tubes (PMTs). To increase the solid angle coverage with detectors with small-size readout system, solutions with Silicon PhotoMultipliers (SiPM) were also studied and implemented, with much effort being made to make the timing performance comparable with the PMT-read detectors. More details on the FAMU X-ray detection system and the related R\&D can be found in Section \ref{sec:scint}.
    
    In addition, the presence of a detector with good energy resolution, even with worse timing performance, is important to always get a check on the X-ray lines present in the spectra and to cross-check the spectra of the scintillators. On this purpose, a coaxial High-Purity Germanium (HPGe) detector has been installed. This detector is described in Section \ref{sec:ge}. 

\begin{figure}[htbp]
        \centering
        \includegraphics[width=\textwidth]{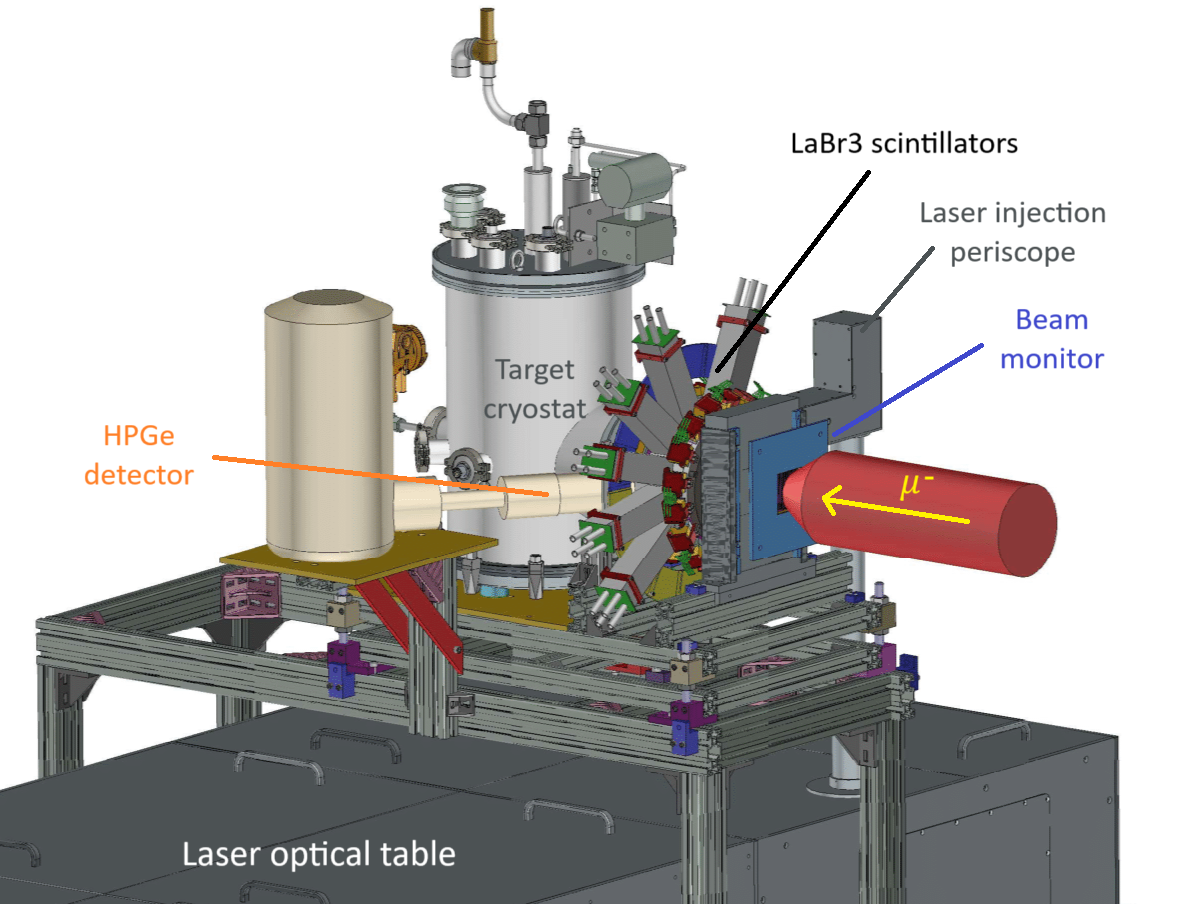}
        \caption{CAD drawing of the 2023 FAMU setup. The target is positioned over the laser optical table, from which the laser is injected in the target through a periscope. The muon beam enters the target after passing through the beam hodoscope (muon beam monitor) and a lead collimator. All 35 X-ray detectors (34 LaBr$_3$ scintillators and 1 HPGe detector) are positioned around the gas target. \label{fig:CAD}}
    \end{figure}

    A sketch of the 2023 FAMU experimental setup is shown in Figure \ref{fig:CAD}, where the laser, target and detectors position are marked.

    \begin{figure}[htbp]
        \centering
        \includegraphics[width=0.8\textwidth]{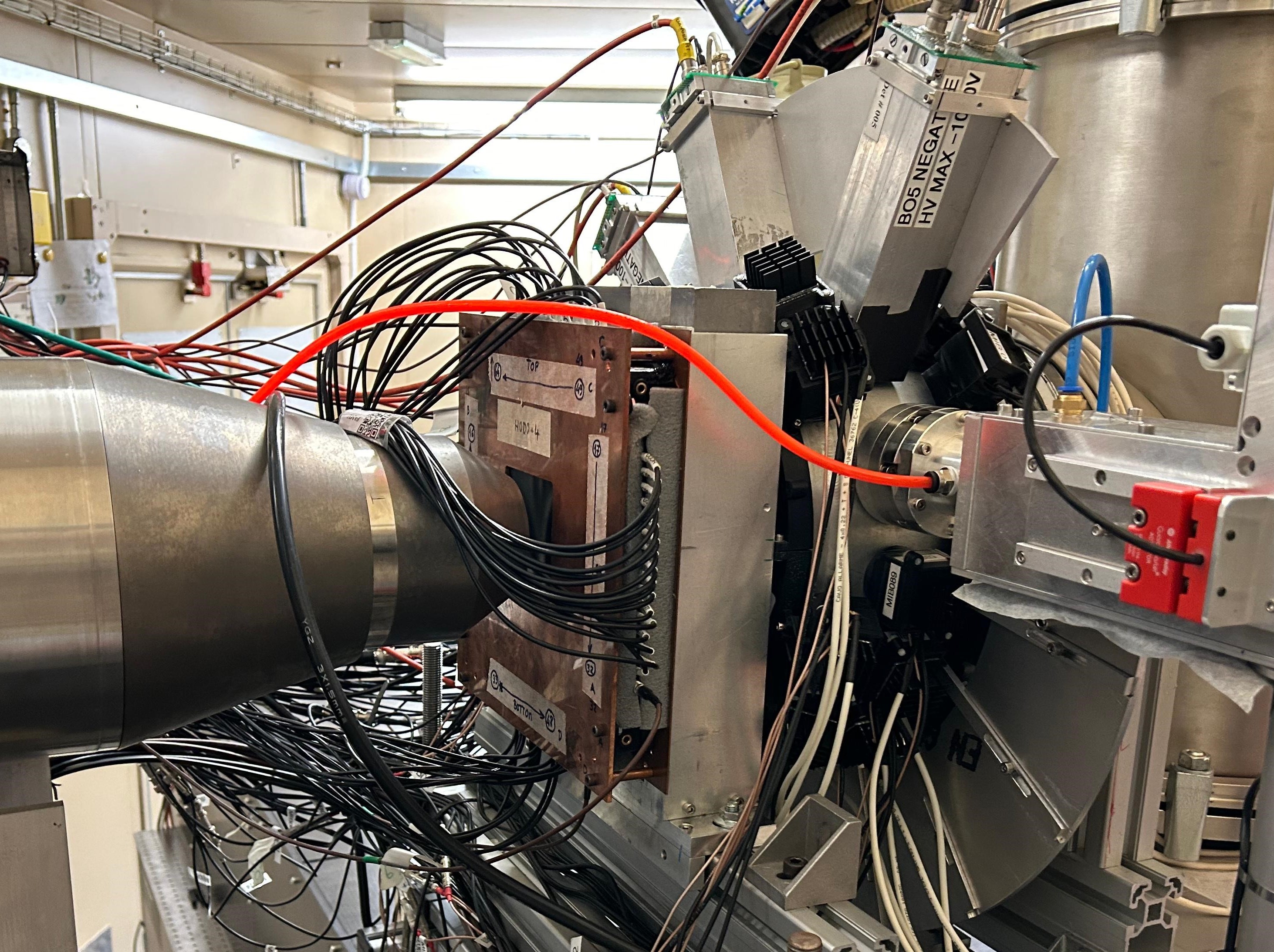}
        \caption{Picture of the 2023 FAMU experiment setup. Left to right: muon beamline collimator, beam hodoscope, FAMU lead collimator, three annular detector holders around the target. The big metal boxes are the detectors read by PMTs, whereas the ones in black plastic-printed cases are those read by SiPMs. The periscope, the metal stucture entering the target from the right side, is also visible.\label{fig:pic}}
    \end{figure}

\section{Muon beamline, laser and target}\label{sec:target}
    A $\upmu^-$ pulsed beam is delivered to the FAMU experiment through the RIKEN-RAL facility\cite{matsuzaki2001}. All three species of pions are produced by impinging the 800 MeV proton beam produced by the ISIS synchrotron against a 1 cm thick graphite target. The decay of negative pions gives rise to a negative muon beam, which undergoes momentum selection and beam focusing in the RIKEN-RAL beamline\cite{hillier2019}. Eventually, the beam is directed against Port1 and reaches the FAMU target. Muon beam momentum and focusing optimisation is crucial to maximise the amount of muonic atoms formed in the target, and therefore to increase the signal-to-noise ratio. The FAMU experiment working momentum is in the range between 50 and 60 MeV/c. The beam is delivered with an average rate of 40 Hz, and each injection consists in two $\sim 50$ ns muon spills separated by $\sim 320$ ns. The average flux with a momentum of 50-60 MeV/c is $ \sim 7 \cdot 10^4$ muons per second.

    A pulsed MIR laser system with tunable wavelength around 6.78 $\upmu$m has been developed for the experiment\cite{baruzzo2023}. The laser beam is produced by combining two laser beams of wavelength 1064 nm and 1262 nm and injecting them into a Difference-Frequency Generation (DFG) crystal. The laser is capable of injecting an energy of $\sim 1$ mJ per pulse, with a rate of 25 Hz. This rate has been set in order to alternate muon spills with/without laser injection, in order to minimise the contribution of systematics due to drifts in the target and detector conditions.  

\begin{figure}[htbp]
        \centering
        \includegraphics[width=\textwidth]{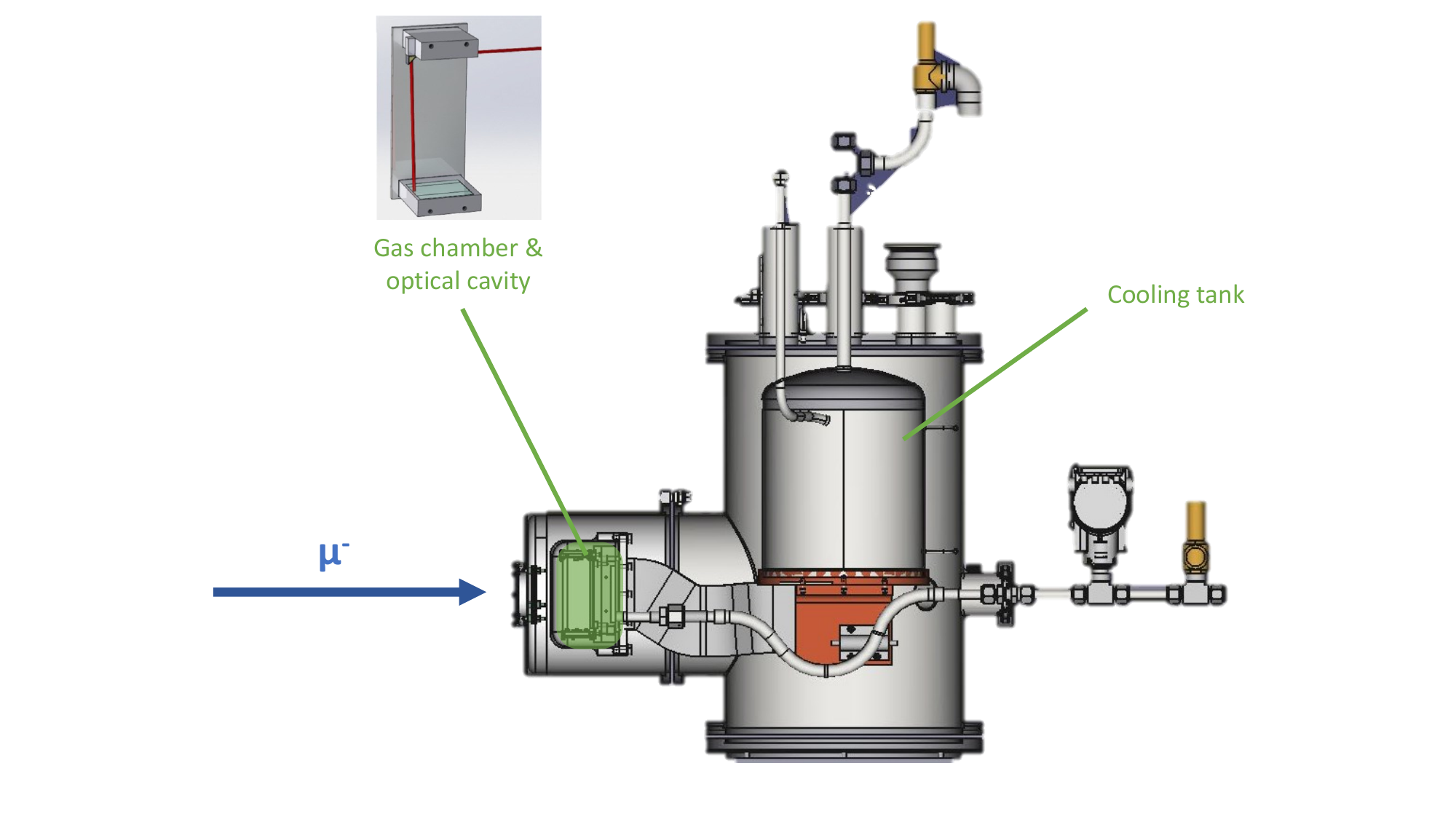}
        \caption{Side section view of the target cryostat (CAD drawing), where the target chamber, the liquid nitrogen tank and the muon beam direction are clearly marked. The three detector-holding rings are attached to the target cryostat around the target chamber (green volume). \label{fig:target}}
    \end{figure}

    The target\cite{pizzolotto2020} consists in a $\sim 10 \times 10 \times 4$ cm$^3$ chamber filled with 7.5 bar of gas mixture (in weight: 98.5\% hydrogen, 1.5\% oxygen) cooled with liquid nitrogen (LN$_2$). The cooling system requires both the target chamber and the LN$_2$ tank to be stored in a cryostat holding $10^{-5}$ bar cryogenic vacuum. A schematic of the target cryostat is shown in Figure \ref{fig:target}. Furthermore, the target chamber contains a $2.2 \times 2.7 \times 10$ cm$^3$ optical cavity aimed at maximising the laser interactions with the gas. A pressure-proof optical window connects the target to the periscope. This window is the only thermal-weak point of the cooling system, and a constant flow of dry air is injected in the periscope to keep the humidity low and avoid condensation on the window and absorption of the MIR light. 

\section{Beam hodoscope}\label{sec:hodo}
    A beam monitor plays a crucial role in allowing data normalisation based on the muon flux impinging the gas target in each beam spill. In fact, the muon flux is proportional to the number of muonic atoms and hence the number of target atoms for the transition.

    The FAMU beam monitor consists of a grid of $32 \times 32$ polystyrene scintillating fibres, coated with BaSO$_4$\cite{bonesini2017, bonesini2018}. Each fibre has squared section, 1 mm pitch, and it is interspaced with the adjacent fibres by 1 mm. The readout is performed by one end of each fibre with a Hamamatsu S13360-1350 SiPM (active area $1.3 \times 1.3$ mm$^2$, pixel pitch 50 $\upmu$m). 

    As the high instantaneous rate of muons (see Section \ref{sec:target}) does not allow single particle detection, a charge deposit calibration using single particles has proved to be crucial to determine the instant muon flux during each beam spill. After an initial test using cosmic rays, the calibration was carried out at the CNAO synchrotron in Pavia (Italy), with a 50 Hz proton beam with kinetic energy 150 MeV. This beam energy was chosen from Monte Carlo simulation in which the energy deposit of negative muons is in the momentum range of interest for FAMU. Further detail on this procedure is presented in \cite{rossini2023_1, rossini2023_2, rossini_submitted}.

\section{Scintillating detectors}\label{sec:scint}
    The FAMU experiment is featured with three different kinds of scintillating detector, placed around the target chamber by three circular holders, in order to exploit the different features of these crystals. All these detectors are based on lanthanum bromide with cerium (LaBr$_3$:Ce) scintillating crystals. 

\begin{figure}[htbp]
        \centering
        \includegraphics[width=\textwidth]{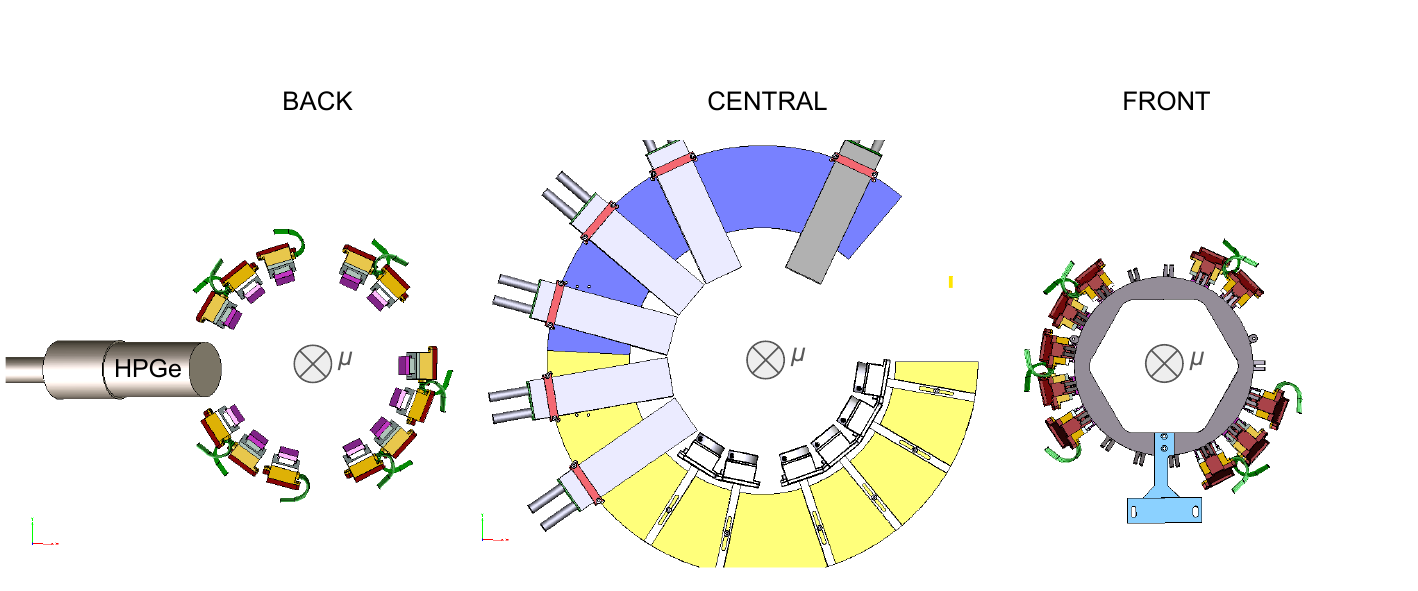}
        \caption{X-ray detector positions in the three detector holders (from the FAMU CAD project). In particular, the front (upstream) ring holds 10 LaBr$_3$ 1" crystals read by SiPMs, the central ring holds 6 LaBr$_3$ 1" crystals read by SiPMs and 6 LaBr$_3$ 1" crystals read by PMTs, whereas the back (downstream) ring holds 12 LaBr$_3$ smaller (1/2") crystals read by SiPMs and it also hosts the HPGe detector. The space left with no detectors on the right side of the central and front rings is occupied by the periscope, as one can see in Figure \ref{fig:CAD}. \label{fig:detectors}}
    \end{figure}

    The first kind of detector featured in the experiment consists in a set of cylindrical crystals, having 1" diameter and 1" thickness, read out by Hamamatsu R11260-200 PMTs, which have a very short rise time ($\sim 12$ ns). Details on these detectors are to be found in \cite{baldazzi2017}. Six detectors of this kind are mounted on the central detector holder, as shown in Figure \ref{fig:detectors} (long gray boxes).

    1" and 1/2" detectors with SiPM readout have also been developed. Crystals for these detectors have two different geometries: cylinders with 1" diameter and 0.5" thickness (called 1" crystals for simplicity), and cubes with 0.5" side (1/2" crystals, for the time being). They use Hamamatsu S14161-3050-AS and S14161-6050-AS SiPM arrays.% at nominal voltages. 
    While 1/2" detectors have a standard parallel ganging, 1" detectors use a custom 4-1 PCB developed with Nuclear Instruments srl, see reference \cite{PCB2023} for more details. For construction details both of 1/2" and 1" detectors see references \cite{bonesini2023} and \cite{bonesini2022}. Twelve 1/2" detectors were mounted on the downstream detector holder, whereas sixteen 1" detectors were positioned in the central (6 detectors) and upstream (10 detectors) rings, as shown in Figure \ref{fig:detectors}. The performances, measured in laboratory, of the 1/2" and 1" detectors are summarised in Table \ref{tab:timeres}\cite{PCB2023, bonesini2023}.

    \begin{table}[htbp]
        \centering
        \caption{Time and energy performance of SiPM-based detectors determined during laboratory tests with radioactive sources, focusing on the 662 keV peak from $^{137}$Cs and the 122 keV peak from $^{57}$Co.\label{tab:timeres}}
        \smallskip
        \begin{tabular}{l|cccc}
            \hline
            & rise time (ns) & fall time (ns) & Res. (\%) @ $^{137}$Cs  & Res. (\%) @ $^{57}$Co\\
            \hline
            1"      & $29.3\pm1.5$  & $147\pm13$    & $3.01\pm0.16$ & $7.9\pm0.4$\\
            1/2"    & $43\pm5$      & $372\pm17$    & $3.27\pm0.11$ & $8.4\pm0.6$\\
            \hline
        \end{tabular}
    \end{table}

\begin{figure}[htbp]
        \centering
        \includegraphics[width=.8\textwidth]{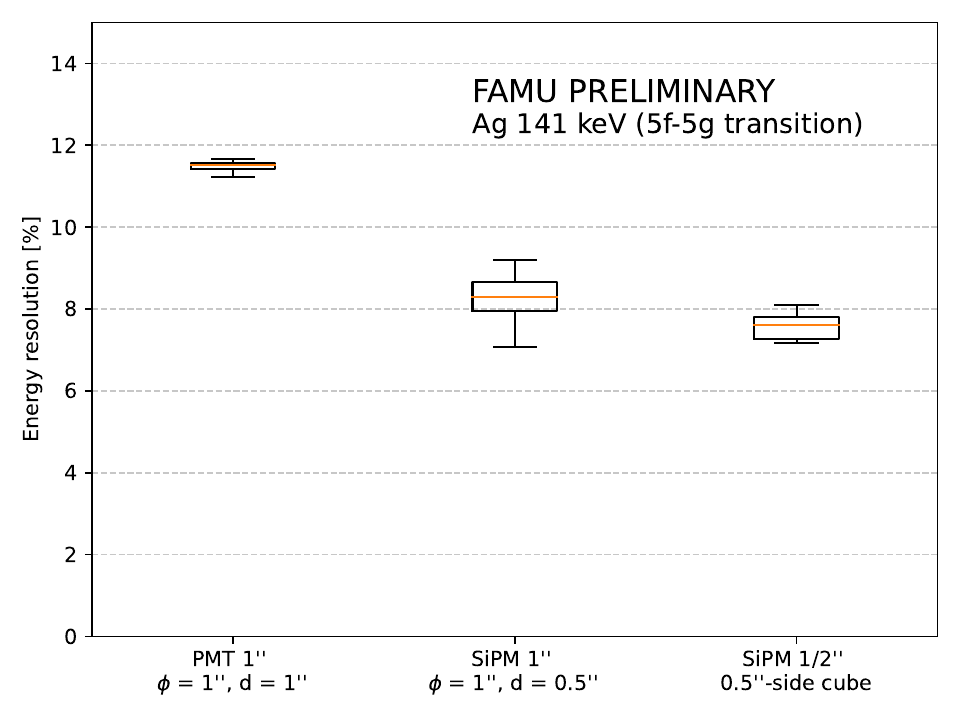}
        \caption{Distribution of FWHM energy resolutions at the muonic silver peak corresponding to the 5f-5g transition (141 keV, theoretically calculated). Preliminary plot: three detectors are still under investigation and were not included; sample size: . The standard boxplot rules hold: a set of box, errorbar and orange line mark the four quartiles in each dataset (i.e. the orange line corrsponds to the median value).\label{fig:enres}}
    \end{figure}

    %  All detectors have been re-calibrated at Port 1 with $^{137}$Cs, $^{241}$Am and $^{133}$Ba sources and the \textbf{linearity plot for a typical detector is shown in Figure 1. (?)}. 
   Energy resolutions on the beamline, obtained from the reconstructed muonic Ag peak at 141 keV for detectors in a typical run on the full experiment setup with 55 MeV/c impinging muons, is reported in Figure \ref{fig:enres}.
    
\section{Germanium detector}\label{sec:ge}
    A commercial ORTEC GEM-S5020P4 High-Purity Germanium (HPGe) Detector has been installed in the FAMU setup along with the scintillating detectors. The detector core is is a p-type coaxial germanium crystal (length 20.8 mm, diameter 30.4 mm) with a 0.9 mm carbon entrance window. The detector is supplied with 1.5 kV positive bias and the signals are treated with an ORTEC A257P preamplifier and an ORTEC 672 gaussian amplifier, with 3 $\upmu$s shaping time.

    From the same measurements as the ones presented in the previous section, the FWHM energy resolution for the HPGe detector at the 141 keV muonic silver peak is $\mathbf{(1.26 \pm 0.17)\%}$. Although the shaping time and the typical signal length of HPGe detectors do not allow to extract timing information in the scale of hundreds of nanoseconds, the good energy resolution is a key factor in the target monitoring and in the beam optimisation for the experiment.

\section{Summary}\label{sec:conc}
    The FAMU experiment has started its data taking at the RIKEN-RAL Port1 $\upmu^-$ beam facility in mid 2023. The X-ray detector setup, presented in detail in this paper, meets the detector requirements using a set of 35 detectors, in particular: 
    \begin{itemize}
        \item 1 High-Purity Germanium detector, mainly for spectra and target check;
        \item 34 LaBr$_3$ scintillators, of which:
        \begin{itemize}
            \item 6 cylindrical 1" diameter $\times$ 1" thickness crystals read by PMTs, all in the central detector holder, which have the best time performance;
            \item 12 cylindrical 0.5" diameter $\times$ 0.5" thickness crystals read by SiPMs, all in the downstream detector holder, with the best energy resolution;
            \item 16 cylindrical 1" diameter $\times$ 0.5" thickness crystals read by SiPMs, all in the downstream detector holder, with the best energy resolution together with the previous ones.
        \end{itemize}
    \end{itemize}

\bibliographystyle{JHEP}
\bibliography{biblio.bib}

%% or
%% [B] Manual formatting (see below)
%% (i) We suggest to always provide author, title and journal data or doi:
%% in short all the informations that clearly identify a document.
%% (ii) please avoid comments such as "For a review'', "For some examples",
%% "and references therein" or move them in the text. In general, please leave only references in the bibliography and move all
%% accessory text in footnotes.
%% (iii) Also, please have only one work for each \bibitem.

%\begin{thebibliography}{99}

% \bibitem{a}
% Author,
% \emph{Title},
% \emph{J. Abbrev.} {\bf vol} (year) pg.

% \bibitem{b}
% Author,
% \emph{Title},
% arxiv:1234.5678.

% \bibitem{c}
% Author,
% \emph{Title},
% Publisher (year).

% \end{thebibliography}
\end{document}